\documentclass[sigconf,anonymous=false]{acmart}

\usepackage{sigirstyle}

\usepackage{amsmath}
\usepackage{booktabs}
\usepackage{tabularx}

\usepackage[utf8]{inputenc}
\usepackage{url}

\newcommand{\topk}{\mathcal{P}_k(\vec{x})}

\setcopyright{acmlicensed}
\copyrightyear{2024}
\acmYear{2024}
\setcopyright{acmlicensed}\acmConference[SIGIR '24]{Proceedings of the 47th International ACM SIGIR Conference on Research and Development in Information Retrieval}{July 14--18, 2024}{Washington, DC, USA}
\acmBooktitle{Proceedings of the 47th International ACM SIGIR Conference on Research and Development in Information Retrieval (SIGIR '24), July 14--18, 2024, Washington, DC, USA}
\acmDOI{10.1145/3626772.3657842}
\acmISBN{979-8-4007-0431-4/24/07}

\begin{document}

\title{``In-Context Learning'' or: How I learned to stop worrying and love ``Applied Information Retrieval''}

\author{Andrew Parry}
\authornote{Author names in alphabetical order by first name.}
\orcid{1234-5678-9012}
\email{a.parry.1@research.gla.ac.uk}
\affiliation{%
  \institution{University of Glasgow}
  \city{Glasgow}
  \country{UK}
}
\author{Debasis Ganguly}
\authornote{Corresponding Author}
\email{Debasis.Ganguly@glasgow.ac.uk}
\orcid{1234-5678-9012}
\authornotemark[1]
\affiliation{%
  \institution{University of Glasgow}
  \city{Glasgow}
  \country{UK}
}
\author{Manish Chandra}
\authornotemark[1]
\email{m.chandra.1@research.gla.ac.uk}
\orcid{1234-5678-9012}
\affiliation{%
  \institution{University of Glasgow}
  \city{Glasgow}
  \country{UK}
}

\renewcommand{\shortauthors}{Parry, Ganguly, and Chandra}

\begin{abstract}
With the increasing ability of large language models (LLMs), in-context learning (ICL) has evolved as a new paradigm for natural language processing (NLP), where instead of fine-tuning the parameters of an LLM specific to a downstream task with labeled examples, a small number of such examples is appended to a prompt instruction for controlling the decoder's generation process. ICL, thus, is conceptually similar to a non-parametric approach, such as $k$-NN, where the prediction for each instance essentially depends on the local topology, i.e., on a localised set of similar instances and their labels (called few-shot examples).
This suggests that a test instance in ICL is analogous to a query in IR, and similar examples in ICL retrieved from a training set relate to a set of documents retrieved from a collection in IR.   
While standard unsupervised ranking models can be used to retrieve these few-shot examples from a training set, the effectiveness of the examples can potentially be improved by re-defining the notion of relevance specific to its utility for the downstream task, i.e., considering an example to be relevant if including it in the prompt instruction leads to a correct prediction. With this task-specific notion of relevance, it is possible to train a supervised ranking model (e.g., a bi-encoder or cross-encoder), which potentially learns to optimally select the few-shot examples. We believe that the recent advances in neural rankers can potentially find a use case for this task of optimally choosing examples for more effective downstream ICL predictions.

\end{abstract}



\received{20 February 2007}
\received[revised]{12 March 2009}
\received[accepted]{5 June 2009}

\maketitle

\begin{figure}[t]
    \centering
    \includegraphics[width=.75\columnwidth]{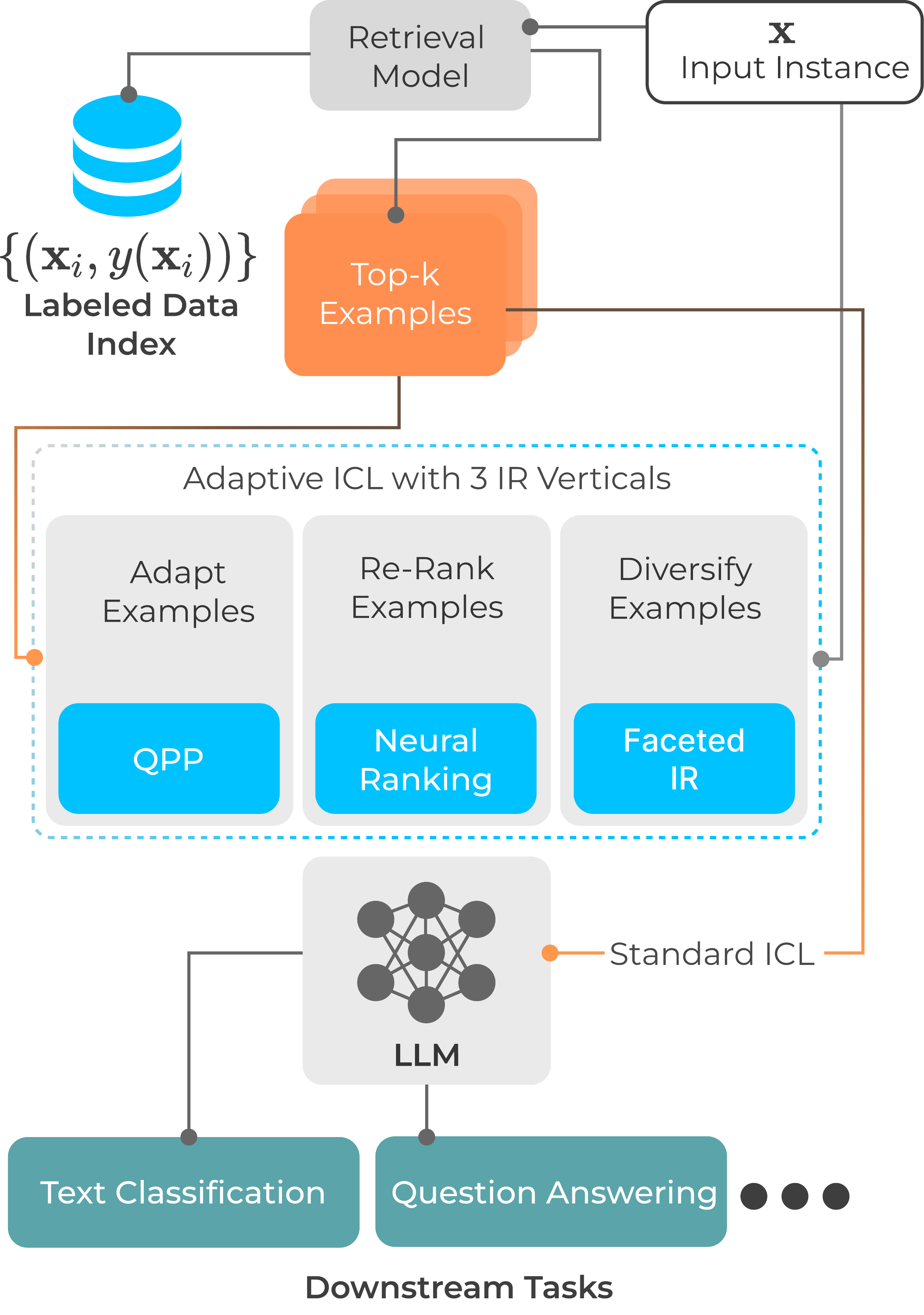}
    \caption{A workflow diagram illustrating how three verticals of IR research fit into the workflow of in-context learning (ICL). Section \ref{sec:aicl} discusses possible ways of adjusting unsupervised and supervised QPP approaches for adapting the number of ICL examples. Section \ref{sec:sicl} discusses ideas of how to learn the notion of downstream usefulness of examples. Section \ref{sec:dicl} discusses methodologies related to diversifying examples for ICL.}
    \label{fig:perspective}
\end{figure}

\section{Introduction} \label{sec:vision}

Research on large language models \cite{openai2023gpt4} (LLMs) is expanding in scope and yielding significant scientific advancements rapidly. These language models are pre-trained on large corpora of documents %
to capture the inherent semantics of text in a generic task-independent manner. Common pre-training methodologies either involve a masked language model (MLM) that predicts randomly masked tokens from the text \cite{BERT,liu2019,reimers2019sentencebert}, or an auto-regressive model or causal language model (CLM)
which predicts a token only from its predecessor tokens \cite{gpt2,gpt3,wang2022gpt}. While MLM is employed in BERT \cite{BERT} and its successors, such as RoBERTa \cite{liu2019}, BART \cite{bart} etc., the latter class of models, i.e., CLM, is applied to train GPT variants \cite{gpt2,gpt3,openai2023gpt4} and open-source Llama and Mistral variants \cite{llama2, jiang2023mistral} etc. 
LLMs, when scaled from millions to billions of parameters, have demonstrated to be
adaptable to a broad set of tasks 
due to instruction tuning \cite{NEURIPS2022_b1efde53},
in the sense that they are not only able to produce semantically correct and coherent text but are also able to adapt themselves surprisingly well with small changes in contexts supplied as inputs, commonly called prompts \cite{arora2022ask}.

This ability to adapt to unseen data and tasks with only a small number of examples differs from the standard notion of supervised learning, where the parameters of a pre-trained model (e.g., BERT \cite{BERT}) is then again learned (commonly referred to as `fine-tuning') from a training set of labelled examples. Instead, in few-shot learning or \textbf{in-context learning} (\textbf{ICL}), a small number of labelled examples from a training set are simply appended to a prompt instruction to control the text generation in a desirable way beneficial to the downstream task \cite{mysore2023large,li-etal-2022-encoder,ni2021large,pradeep2023does}. In addition to leveraging ICL for a purely generative task, e.g., question answering or abstractive summarisation \cite{gpt3, li-etal-2023-shot, tang-etal-2023-context}, a more common use is in a predictive task, such as text classification \cite{lu-etal-2022-fantastically,milios-etal-2023-context}, where each class is specified by a set of words (commonly called a verbaliser \cite{schick-schutze-2021-exploiting}), e.g., for a binary sentiment classification task the positive class could be defined by the set $\{$`good', `great', `wonderful'\ldots$\}$.
Once each class for a predictive task is well-defined, the generated text can be mapped to the most likely class(es) by using the posterior over the vocabulary generated by the decoder.

It is to be realised that ICL is somewhat conceptually similar to a non-parametric approach, such as $k$-NN, where the prediction for each instance essentially depends on the local topology, i.e., on a localised set of similar instances and their labels (called few-shot examples) - the only difference of ICL with $k$-NN is that the former involves a frozen set of encoder-decoder parameters of an underlying LLM, where ICL generally works well on any domain with only a small number of examples because, unlike supervised models, it does not suffer from overfitting the parameters on a particular set of labelled examples. Instead, the semantics expressed in the examples potentially play a key role in controlling the text generation process to yield the desired output—text generation process to yield the desired output -- either the text itself or a mapping into a class prediction.

The usefulness of localised examples, akin to the nearest neighbor-based prediction, suggests a strong case of analogy between ICL and ad-hoc IR. More precisely, a test instance in ICL is analogous to a query in IR, and similar examples in ICL retrieved from a training set relate to a set of documents retrieved from a collection in IR. This analogy opens up several interesting research questions in ICL concerning the effective use of IR to improve ICL predictions. In this perspective paper, we discuss particular components of ICL that can be mapped to known and well-researched IR problems. This means that the solutions to these problems that the IR community has researched over decades can potentially be applied to improve the effectiveness of ICL. Moreover, this should also be of interest to an IR researcher to develop new methodologies for classic problems in IR, such as that of document retrieval or query performance prediction (QPP) \cite{deep_qpp, qpp_multifield, unsupervised_qpp}, specifically catered to a downstream predictive task, hence opening up new possibilities of evaluating novel IR methodologies beyond retrieval tasks.

We now propose the following three main ways of incorporating the ideas from core IR into ICL. First, during inference in ICL, instead of using a constant number of examples for each instance, a potentially better approach could be to make the \textbf{number of examples variable}.
A similar problem in IR is to predict how many documents to retrieve (or equivalently to predict the rank cutoff threshold \cite{stop_reading,choppy}), which is also closely tied to the problem of \textbf{query performance prediction} (QPP) \cite{deep_qpp, qpp_multifield, unsupervised_qpp}. In the context of ICL, this means that for some test instances, one can find more `useful' examples from the training set (an example can be considered useful if including it as a part of the prompt leads an LLM to the correct prediction). In contrast, it is difficult for others to find such useful examples. An ICL method that is aware of the quality of the examples can, hence, potentially adapt itself, e.g., by using a higher number of examples if the example quality is predicted to be poor.    

Second, we propose to make the underlying metric space -- used to compute the similarities between a test instance and the examples -- learnable. The objective of learning this similarity function would be to rank the `useful' examples ahead of the not useful ones. Unsupervised retrieval models only consider the similarity between the textual content of a test instance and the training instances. However, a \textbf{supervised retrieval model} learned via a standard ranking objective (e.g., noise-contrastive loss~\cite{nce}) using triplets -- each triplet constituting a test instance (query), a useful example (relevant document), and a not useful one (non-relevant document) -- may specifically capture the inherent semantics of the utility of examples for a particular test instance.

Third, \textbf{diversity of examples} is likely to play a part in the effectiveness of ICL because an example that is different from the previous one selected should be more informative to an LLM decoder to generate relevant words which can then be mapped to the correct class. This can also be traced to facet or aspect-based IR which attempts to make the top retrieved set of documents cater to all the latent aspects of an information need \cite{upadhyay2020aspect, faceted_search}.

These three core tasks in IR, namely that of \textbf{QPP}, supervised ranking or \textbf{learning to rank}, and \textbf{diverse or faceted IR} have a long history of thorough investigation, which have constantly pushed the boundaries of the state-of-the-art achievable for these tasks. In this perspective paper, we argue that this knowledge gained by the IR community could be beneficial to further improve the effectiveness of generative AI for text.

In the next section (Section \ref{sec:icl}), we provide a brief technical introduction to the concept of ICL, following which, we structure the remainder of the paper into three sections detailing on how each of the specific IR tasks can be applied to an ICL workflow, i.e., QPP for adaptive ICL (Section \ref{sec:aicl}), learning to rank for learning to order examples in ICL (Section \ref{sec:sicl}), and diversity-based and faceted IR for obtaining more informative examples in ICL (Section \ref{sec:dicl}). Although an exhaustive empirical validation of each of these independent ideas for improving ICL is beyond the scope of a perspective paper, we do, however, include a preliminary evaluation to support the use-case of QPP in ICL (Section \ref{sec:eval}), where we show that adjusting the number of examples in a data-driven manner does lead to significant improvements. We believe that this focused investigation, along with the other ideas presented, would motivate other NLP researchers to apply black-box established IR methodologies or even IR researchers to adjust the state-of-the-art IR methodologies to specifically cater to the downstream predictive tasks in ICL.

\section{In-Context Learning}  \label{sec:icl}

We first provide a brief technical introduction to In-Context Learning (ICL) before describing how the ICL methodology can be improved by incorporating core ideas from IR.

\subsection{A Formal Introduction}
In-context learning (ICL), unlike supervised learning, does not involve training a set of parameters $\theta$ on labeled examples. Rather, the posteriors are now a function of the following: a) text of the input test instance, b) the decoder parameters of a pre-trained large language model (LLM), c) a prompt instruction, and d) optionally, a set of $k$ input examples (commonly called $k$-shot learning). Formally,
\begin{equation}
P(y|\vec{x}) = f(\vec{x}, \mathcal{P}_k(\vec{x}); \phi_{\text{LLM}}),
\label{eq:icl}
\end{equation}
where, different from a supervised setup, the function $f$ does not have a parameterized representation that can be learned using a training set with gradient descent. The function itself depends on the pre-trained parameters $\phi_{\text{LLM}}$ of an LLM, the current inputs for which a label is to be predicted, and a prompt comprising a set of $k$ text units denoted by $\mathcal{P}_k(\vec{x})$.

\begin{figure*}[t]
\centering
\includegraphics[width=1.5\columnwidth]{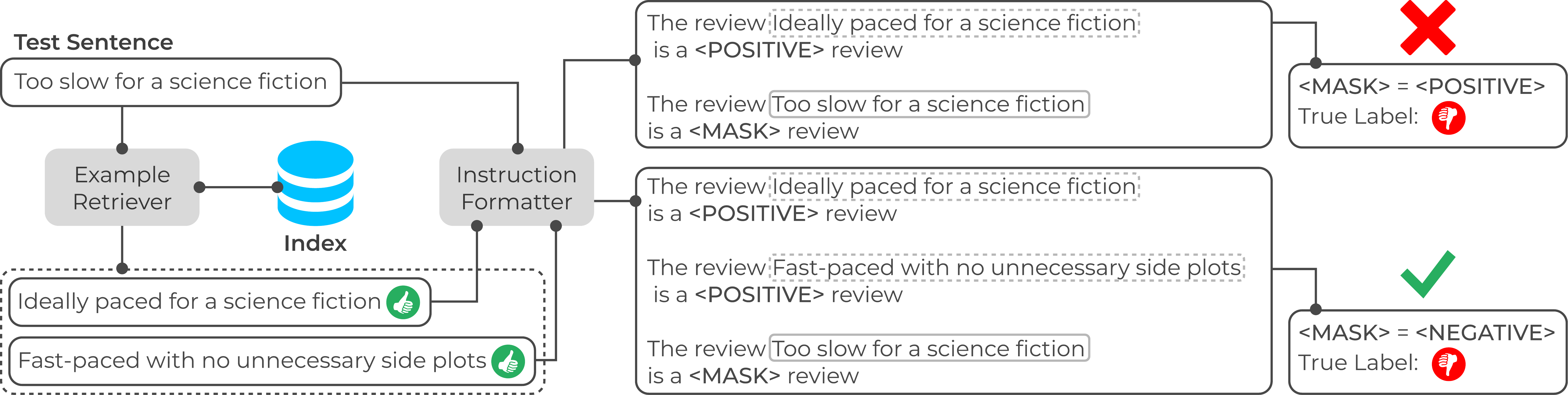}
\caption{Example workflow of In-Context Learning for sentiment classification. The illustrative example shows a sample test instance for which a single demonstration (as retrieved from the training set) does not result in the correct prediction (prediction shown at the top). The example also shows that increasing the number of demonstrations from one to two results in the correct prediction (shown at the bottom). Demonstrations included within the prompt are shown in blue. 
\label{fig:icl-workflow}
}
\end{figure*}

Since the decoder of an LLM generates a sequence of words of the form of $w_1,\ldots,w_N$ ($N$ being the maximum length of a sequence), the class posterior likelihoods are computed in the following way. A set of classes (say for a $p$-way classification problem) is mapped to $p$ different equivalent sets of words, say $V(y)$, where $y \in \mathbb{Z}_p$ -- these sets commonly being called verbalisers \cite{hu2021knowledgeable}. For instance, for a binary classification problem (e.g., that of a movie review as shown in Figure \ref{fig:icl-workflow}), $p=2$ (i.e., $y \in \{0, 1\}$), and a reasonable way to define the verbaliser sets could be via the following words: $V(0) = \{\text{`false'}, \text{`negative'}\}$, and $V(1) = \{\text{`true'}, \text{`positive'}\}$.

Note that the word `learning' in ICL is a misnomer because there are no updates to the decoder parameters of an LLM. For more details on ICL, please refer to these excellent surveys \cite{dong2023survey,luo2024incontext}. 

\subsection{The role of IR} \label{ss:roleofir}
One of the most important components of ICL (as shown in Figure \ref{fig:icl-workflow}) is the search component that outputs a top-$k$ candidate set of \textit{similar} instances from the training set, i.e., $\topk$ of Equation \ref{eq:icl}. Although, in principle, it is possible to include random examples from the training set in the prompt, it has been shown that localised examples (i.e., examples that are topically similar to the current instance) yield better performance \cite{liu-etal-2022-makes,luo2024incontext}. The reason why this works can be traced to the fundamental principle of reproducing kernel Hilbert spaces (RKHS) machine learning -- that a predictor function is an aggregation of parameterised kernel functions pivoted around training data instances \cite{Paulsen_Raghupathi_2016}.

It is thus crucial to retrieve as many relevant examples as possible from the training set while imposing a practical constraint on the number of such examples for efficiency reasons -- a classic trade-off of recall and precision in IR ad-hoc retrieval;
the only difference is that relevance for ICL needs to be defined in terms of the utility or usefulness of an example towards the correct prediction.  

A similar question explored in IR is where to stop reading a ranked list because there is little utility in retrieving documents due to the low probability of finding relevant documents beyond a certain rank cutoff \cite{stop_reading,choppy}. What is more challenging is that this rank cut-off depends on the number of relevant documents occurring in the collection for a specific query, that is to say, while some queries with well-defined information needs are associated with a small number of relevant documents satisfying the specific relevance criterion, other queries with broader information needs usually are associated with a higher number of relevant documents \cite{CarteretteKHC14}. In core IR research, this problem is usually addressed by estimating the retrieval qualities of queries -- the assumption being that well-specified queries yield better retrieval results (in terms of precision and recall), whereas ill-specified ones suffer from poor retrieval quality due to the apparent ambiguity of information need. This motivation paves the path to the following section, where we discuss how query performance prediction (QPP) can also be beneficial to the related problem of retrieving similar examples in ICL.

\section{Adaptive ICL $\mapsto$ QPP?} \label{sec:aicl}

In this section, we describe an adaptive approach to the selection of examples for ICL. We outline analogous principles from IR literature that can be applied in broader tasks.

\subsection{A Variable Number of Examples}

\begin{figure}[t]
\centering
\includegraphics[width=0.65\columnwidth]{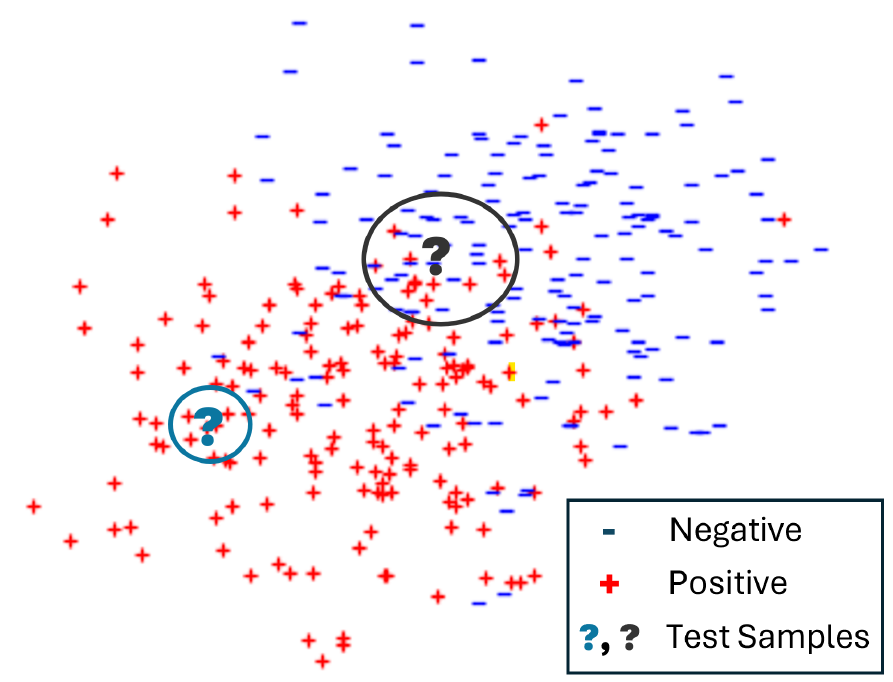}
\caption{
Motivation behind using a variable sized neighborhood for $k$-NN classification \cite{learning_k_knn}: An instance close to a decision boundary (black `?') is likely to have a higher heterogeneity in its class distribution, thus indicating the necessity of a larger neighborhood for an effective classification.
}
\label{fig:k_in_knn}
\end{figure}

The observation in IR that different queries exhibit different levels of retrieval performance can be utilised for ICL, where we can draw an analogy that some test instances are associated with better candidates for training examples (i.e., examples which are useful in the sense that including them as a part of the prompt leads to correct predictions), and hence including a small number of them should be adequate. On the other hand, the retrieval quality for some test instances (used as queries for ICL) does not yield good candidates. As a result, one needs to look down the ranked list further to collect useful examples.

We call this methodology of using a variable number of demonstrations for ICL inference by the name `\textbf{Adaptive In-Context Learning}', or AICL for short.
The idea of AICL centres around choosing the context $\mathcal{P}_k(\vec{x})$ in a data-driven manner, i.e., making $k$ a function of the data (current instance $\vec{x}$) itself. This is somewhat analogous to choosing different values of $k$ for a $k$-NN based non-parametric modeling \cite{learning_k_knn}, as shown in Figure~\ref{fig:k_in_knn}. The motivation is that classifying some instances would be more difficult than others, in which cases they are potentially to be benefited from a larger value of $k$ (more context). On the other hand, for relatively easy data instances using too much context may be detrimental for effective prediction.

Formally speaking, the difference of AICL with that of ICL (Equation \ref{eq:icl}) is that the value $k$, indicating the size of the neighborhood, is no longer a constant. Instead, we denote it by a parameterised function $\kappa(\vec{x})$ such that
\begin{equation}
P(y|\vec{x}) = f(\vec{x}, \mathcal{P}_{\kappa(\vec{x})}(\vec{x}); \phi_{\text{LLM}}),
\label{eq:aicl}
\end{equation}
where $\kappa: \vec{x} \mapsto \{0,\ldots,M\}$,
$M$ being an upper bound on the number of example instances.
We now suggest how unsupervised or supervised approaches may be applied to choose the rank cutoff $\kappa$.

\subsection{Unsupervised Rank Cutoff} \label{ss:unsup-rankcut}
Among unsupervised approaches, two main ideas in IR research can be used to determine the number of examples in ICL.

\para{Score Distribution-based Models} The first thread of work is based on the hypothesis that the scores of relevant and non-relevant documents follow a different statistical distribution, e.g., \citeauthor{stop_reading} propose to use a mixture of Normal-Exponential distributions -- Normal for relevant and Exponential for non-relevant documents -- to model the score distribution of top-ranked documents. The work in \cite{stop_reading} uses expectation maximisation (EM) to estimate the parameters of this mixture distribution and thereby predict the most likely cutoff rank beyond which the probability of finding a relevant document is considerably low. Such ideas of utilising the characteristic differences between the score distributions of relevant and non-relevant documents have also been used for query performance prediction (QPP) \cite{cummins}.

While an EM from retrieval scores allows provision for applying a variable number of examples, the following are some ICL-specific challenges that need to be researched.
\uls
\li With the notion of relevance being changed to \textbf{`downstream utility'}, the score distributions of useful and not useful examples may not follow the same mixture distribution of Normal-Exponential as reported in \cite{stop_reading,cummins}. It will be an interesting future research direction to investigate the latent relations between the similarity scores and the downstream utility of the examples in the context of ICL.

\li With a threshold on the score-distributions, it is difficult to restrict the cutoff to a maximum value, which is essential for ICL due to a maximum limit on the input size to an LLM.

\li A score distribution-based approach does not explicitly consider the information from the queries themselves (equivalently, the test instances in ICL).
\ule
We now describe another thread of work in IR research that may help alleviate the last two limitations. 

\para{QPP-based Models}

Different from the rank cut-off strategies, query performance prediction (QPP) models seek to estimate the retrieval quality of a query. As a direct analogy, such methods can be applied to the top-similar examples retrieved in ICL with a different objective of predicting the usefulness of the examples.

Most of the classic works in QPP involve unsupervised approaches that make use of the information from the set of top-retrieved documents to estimate how topically distinct are the top-retrieved documents from the rest of the collection -- a large difference indicating potentially better retrieval quality \cite{croft_qpp_sigir02}. Various evidences extracted from the top-retrieved documents have been shown to be useful for different post-retrieval QPP estimation methods. This includes i) the KL divergence between the language model of the top-retrieved documents and the collection model in Clarity \cite{croft_qpp_sigir02}, ii) the aggregated values of the information gains of each top-retrieved document with respect to the collection in WIG (Weighted Information Gain) \cite{wig_croft_SIGIR07}, iii) the skew of the RSVs (Retrieval Status Values) measured with variance in NQC (Normalized Query Commitment) \cite{kurland_tois12}, iv) ideas based on the clustering hypothesis for a pairwise document similarity matrix \cite{fernando_correlation_sigir07}, and, more recently, v) the characteristics of the embedded space of documents and queries \cite{qpp_DG_IPM,FaggioliFerroEtAl2023}.

A suitable adaptation of these existing techniques can be applied in a two-stage pipeline to determine the number of examples in ICL. As a first step, one can employ a QPP methodology to predict the retrieval quality (in terms of the usefulness) of a set of ordered examples -- a high value likely indicating that the useful examples can potentially be found at the very top ranks, as a result of which, a small number of examples should potentially work well. On the other hand, a low QPP estimate likely indicates that the very top ranked examples are not likely to be useful for downstream prediction, in which case it should be better to employ a large number of examples.
This approach of selecting rank cutoffs (with an upper bound) as a function of the QPP scores has been applied to determine a variable depth of relevance assessments required for a robust retrieval evaluation \cite{DBLP:conf/sigir/GangulyY23}.

\subsection{Supervised Rank Cutoff} \label{ss:sup-rankcut}

Instead of devising a heuristic to predict the number of training examples to use for a test instance $\vec{x}$, i.e., $\kappa(\vec{x})$, a supervised approach can be applied to solve this as a classification problem, i.e., $\kappa \equiv \text{Softmax}(\vec{x}^\mathrm{T}\theta)$, %
where $\theta$ is a set of layer(s) of parameters. The underlying hypothesis is that if we provide enough training data constituting the optimal number of examples for a range of topical content, we should be able to learn to predict the likely number of examples to use for unseen text during inference time.

To train a classifier that maps a text to a number between 1 to $M$ (the maximum number of examples), it is necessary to obtain the ground-truth labels, i.e., the optimal number of examples, for each instance in the training set. We propose to obtain this by the following methodology:
Given a training set instance $\vec{x}$, one can employ a similarity function (e.g., BM25) to retrieve a candidate set of $M$ examples - $\{\vec{z}_1,\ldots,\vec{z}_M\}$. Since $\vec{x}$ is an instance from the training set, we can utilise its label to check if the $k$-shot predictions using an LLM are correct. It may happen that correct predictions are obtained for several values of $k \in \{1,\ldots,M\}$. Several strategies can be adapted to define the ground-truth number of examples. For instance, one can stop early and simply select the smallest $k$ that results in a correct prediction. Alternatively, a potentially more robust procedure would be to exhaustively check through all possible values of $k=1,\ldots,M$, and select the one that results in a correct prediction with the least uncertainty \cite{rubin-etal-2022-learning,sorensen-etal-2022-information}.

The workflow of this least uncertainty-based selection of the ground truth for the number of ICL examples is shown in Algorithm \ref{algo:buildgt}. Algorithm \ref{algo:getpreds}, which is invoked during the ground-truth construction, shows a sample prompt template for text classification.

\begin{algorithm}[t]
\small
\DontPrintSemicolon
\KwIn{$\vec{x}$ -- an instance from the training set}
\KwIn{$k (< M)$ -- number of examples (max $M$)}
\KwOut{$\Delta_p$ -- Softmax posteriors}
\Begin{
$N_k(\vec{x}) \gets \{\vec{z}_1,\ldots,\vec{z}_k\}$ 

\texttt{Instruction} $\gets$ 
``Predict the type of $\langle \vec{x} \rangle$ as one of $\{ \langle C_0 \rangle, \ldots, \langle C_{p-1} \rangle\}$ given the following example''.\\
\For{$i \gets 1$ to $k$}{
\texttt{Instruction.append(}``Example: $\langle \vec{z}_i \rangle$ is a representative of class $\langle y(\vec{z}_i) \rangle$''\texttt{)}\\
}
$\Delta_p \gets$ \texttt{LLM(Instruction)}\\
\KwRet {$\Delta_p$}
}
\caption{LLM $k$-shot predictions}
\label{algo:getpreds}
\end{algorithm}

\begin{algorithm}[t]
\small
\DontPrintSemicolon
\KwIn{$\mathcal{T}$ -- a training set of labelled instances}
\KwOut{$\mathcal{K} = \cup_{\vec{x} \in \mathcal{T}}\optk(\vec{x})$ -- Number of examples yielding the most confident and correct predictions for each instance $\vec{x} \in \mathcal{T}$}
\Begin{

\For{$\vec{x} \in \mathcal{T}$}{

\texttt{max\_confidence $\gets$ 0; $\optk \gets 1$}

\For{$j \gets 0$ to $M$}{
$\Delta_p \gets$ \texttt{LLM $k$-shot predictions($\vec{x}, j$)} \tcp*[r]{\footnotesize Call Algorithm \ref{algo:getpreds}, i.e., try to predict with $j$ examples}

$\hat{y}(\vec{x}) \gets \text{argmax}\Delta_p$ \tcp*[r]{\footnotesize Get the predicted class}
\texttt{confidence}$\gets \Delta_{\hat{y}(\vec{x})} \mathbb{I}(\hat{y}(\vec{x}) = y(\vec{x}))$ \tcp*[r]{\footnotesize Check if the predicted class is the correct one and record the prediction confidence}

\If {\texttt{confidence $>$ max\_confidence}}{
\texttt{max\_confidence $\gets$ confidence} \tcp*[r]{\footnotesize Keep track of the least uncertain correct prediction}
$\optk \gets j$
}
}
$\mathcal{K} \gets \mathcal{K} \cup \optk$
}
\KwRet {$\mathcal{K}$}
}
\caption{Optimal number of examples}
\label{algo:buildgt}
\end{algorithm}

After executing Algorithm \ref{algo:buildgt}, we obtain a set of ground-truth labels $\mathcal{K}$ which could then be used to train a classifier, parameterised by $\theta$, via optimising:
\begin{equation}
\text{argmin}_{\theta} \sum_{
\vec{x} \in \mathcal{T}, \optk \in \mathcal{K}}\mathcal{L}(\vec{x}^{\mathrm{T}}\theta, \optk),    \label{eq:celoss}
\end{equation}
where $\mathcal{L}$ is a standard loss function, e.g., the cross-entropy.

During inference, for each $\vec{x} \in \mathcal{E}$ ($\mathcal{E}$ denoting an evaluation set), we propose to apply the classifier $\kappa: \vec{x} \mapsto \{1,\ldots,M\}$ -- trained via Equation \ref{eq:celoss} -- to predict the number of examples, and eventually conduct a $\kappa(\vec{x})$-shot prediction on $\vec{x}$ (Equation \ref{eq:aicl}).

\subsection{Open Research Questions and Challenges}

Till now in this section, we described how unsupervised and supervised approaches can be applied to dynamically select the number of examples to be used for an ICL-based prediction.
In this section, we discuss some research directions that could be explored to adapt ICL in alternative ways to further improve its effectiveness.

First, we would like to point out to the existing work on generating query variants, as a part of a data augmentation strategy, to devise alternative formulations of the same or similar information needs. This has been shown to improve the effectiveness of rankers \cite{DBLP:conf/acl/GaoMLC23}, query performance prediction \cite{query_variants_kurland,DBLP:journals/tois/DattaGMG23} relevance feedback \cite{DBLP:conf/cikm/ChakrabortyGC20}, and even act as a tool to measure consistency of IR models \cite{DBLP:conf/cikm/Sen0GVR22}. Given the recent success of zero-shot query generation capabilities of LLMs \cite{querygen-llm,wang-etal-2023-query2doc}, we believe that augmenting a test instance with alternative text representations can be useful to eventually improve retrieval quality (and hence potentially improve the downstream ICL effectiveness). The unsupervised and supervised approaches for predicting the number of examples per query (test instance) may also lead to better ICL effectiveness, as per the existing findings that variants do actually help improve QPP \cite{query_variants_kurland,DBLP:journals/tois/DattaGMG23}.
We thus formulate the following two research questions aligned along this direction.
\uls
\li \textbf{RQ-\thesection .1}: Can query variants generated by LLMs (or otherwise) improve the prediction of the number of examples to use for each instance?

\li \textbf{RQ-\thesection .2}: Can relevance feedback based approaches with or without the use of generated query variants help reorder the top-$k$ initially retrieved candidate set of examples towards a better prediction of the number of examples?

\ule

The other direction of work involves a dynamic selection of not just the neighborhood size but also other ICL parameters. For instance, the verbaliser \cite{schick-schutze-2021-exploiting} sets can be selected dynamically from a set of alternatives based on the input instance. Further, a prompt can also be selected dynamically - again based on the input instance; an unsupervised approach exploring this idea has already been studied in \cite{sorensen-etal-2022-information}. Generally speaking, the research question that can potentially be explored is the following.

\uls
\li \textbf{RQ-\thesection .3}: Can other ICL parameters also be chosen in a data-driven manner to lead to better effectiveness, e.g., the verbaliser, the prompt, or even an LLM itself (akin to a mixture of experts)?
\ule

\section{Rank ICL Examples $\mapsto$ Supervised IR?} \label{sec:sicl}

In this section, we discuss another crucial aspect of ICL that can potentially be improved by developing ranking models specifically suited to a different notion of relevance: \textit{ICL downstream task-specific usefulness of examples}. The concept of an effective example in core neural IR is well-researched, particularly the notion of `hard' negatives during fine-tuning~\cite{DPR, infoNCE}. These negatives have improved downstream precision on ranking tasks~\cite{retromae} and, more generally, representation learning~\cite{simcse}.

Specific to few-shot learning, \citet{rubin-etal-2022-learning} employed a noise contrastive estimation (NCE) loss~\cite{nce} to train a bi-encoder-based pairwise ranker using SBERT \cite{reimers2019sentencebert} embeddings.  For training the ranking model, pairs of instances (relevant and non-relevant examples) were collected in the following way. For each pivot instance $\vec{x}$ from a training set, the authors employed BM25 to constitute a top-$k$ candidate set of examples. Each pair $(\vec{x}, \vec{z}_i)$ was then tested to check whether a 1-shot prediction with $\vec{z}_i$ was correct, in which case, $\vec{z}_i$ was classified as a relevant example for $\vec{x}$, or else it was considered as a non-relevant one. Batches comprising relevant and non-relevant pairs were then constituted to train a standard NCE loss.  While the work of \citet{rubin-etal-2022-learning} is a definitive step towards leveraging a task-specific notion of relevance, the investigation should not be considered complete. Several potentially promising research directions should be explored to improve ICL effectiveness further. We now provide a survey of neural ranking literature introducing core paradigms which may be utilised in example selection.

\para{Bi-Encoder architecture}
A bi-encoder architecture encodes text into a latent representation that can be compared in a vector space; in the context of a retrieval task, these texts would be queries and documents. While a bi-encoder is implemented either with a Siamese network of shared parameters~\cite{reimers2019sentencebert} or as a single encoder~\cite{cedr}, the latter has become prevalent in recent years~\cite{DPR, retromae}.

The performance of neural models in search was significantly improved with the release of BERT~\cite{devlin-etal-2019-bert}. \citet{DPR} first proposed the use of `hard' negatives mined from BM25 to improve the precision of BERT-based rankers. \citet{infoNCE} then proposed a variant of the NCE objective, `Localised Contrastive Estimation', in which multiple negatives are sampled for each query to account for the variance in the notion of non-relevance. In doing so, they also showed the effectiveness of hard negatives mined from fine-tuned rankers.
To further improve the quality of negative samples, \citet{ance} proposed that a model could choose negatives during training to allow negatives to become continuously `harder' as fine-tuning progresses.

At a conceptual level, bi-encoders generally represent a text as a single embedding by using the representation of the BERT [CLS] token as a proxy for the entire sequence. Other pooling methods are effective, including maximum sequence similarity~\cite{maxp} and late interaction in which a max pooling is performed over the token-level similarity of each query token to document tokens~\cite{colbert}. More recent works instead use a BERT-style encoder with a shallow decoder, which places greater emphasis on the ability of the encoder during pre-training. This architectural development has yielded not only state-of-the-art recall but new pre-training styles, including lexical grounding~\cite{lexmae} and text reconstruction~\cite{retromae}.

The separate encoding of queries and documents allows for the offline encoding of documents which can vastly improve online latency. This is often coupled with an approximate nearest neighbour search in a vector space~\cite{colbert, crossarchitecture}. More specifically, after training a bi-encoder model, the parameters of the trained model act as `embeddings' for each document in the collection. During inference time, a query is first embedded into a vector. Then an approximate nearest neighbour search, e.g., HNSW~\cite{hnsw}, is conducted on an indexed representation of these dense document vectors. Therefore, exploring the potential benefits gained from efficient, dense end-to-end retrieval of training examples for effective ICL can be an interesting research direction.

\para{Cross-Encoder architecture}
A cross-encoder instead jointly encodes a query and document at inference time~\cite{monobert}, allowing deep interactions between texts that are impossible in a bi-encoder architecture. Empirically, these models are more precise than bi-encoders at the expense of latency, as representations cannot be pre-computed in a standard setting. Both BERT- and T5-based architectures have been proposed~\cite{monobert, monot5}; in the case of a BERT model, a feed-forward classification head is used to output class probabilities of relevance~\cite{monobert}. In the case of a sequence-to-sequence model, token logits are taken as surrogates for class probabilities~\cite{monot5}. Recent developments in LLMs have prompted research in these large decoder-only models as text rankers. A list-wise approach is commonly taken in which a model receives multiple documents for a given query and outputs a permutation of the original ranking~\cite{rankgpt, rankzephyr}. The development of these models is still in its infancy but it offers opportunities to investigate highly precise ranking models potentially in sample mining beyond simple ad-hoc search.

Therefore, it is reasonable to assume that employing a cross-encoder to learn ranking examples by their downstream usefulness should yield better results than a bi-encoder-based approach. An interesting research direction would hence be to investigate the optimal architecture within an ICL pipeline, considering the efficiency-effectiveness trade-off.

\para{Teacher Distillation}
Moreover, a rich literature exists on distilling the more computationally expensive cross-encoder models into the simpler bi-encoder, the former acting as a teacher model and the latter as a student~\cite{crossarchitecture}. Distilling a teacher model into a bi-encoder one allows provision for end-to-end dense retrieval without requiring any sparse index to retrieve a candidate top-$k$. Two core paradigms of distillation are homogeneous architecture and heterogeneous architecture distillation. The former generally will distill one model into a newly initialised copy via minimisation of a divergence metric over either the final hidden state~\cite{TCT} or internal states such as attention layers~\cite{atlas}. The latter minimises prediction error between teacher and student models commonly via a mean squared error criterion over triplet residuals (residual between positive and negative example scores), allowing for `cross-architecture knowledge distillation'~\cite{crossarchitecture} as a scalar relevance score is not architecture dependent. This approach has become a core component of many state-of-the-art dense retrieval models, frequently beginning with a cross-encoder teacher used to mine hard negatives and teacher scores before a second stage distillation is performed using the previous distilled model as a teacher~\cite{retromae, lexmae}. A parallel area of work gaining traction is knowledge sharing between a retrieval system and a generative model~\cite{rag, fid, atlas}. This paradigm is directly correlated with our perspective with recent work finding success in directly optimising a retriever to maximise downstream QA performance~\cite{atlas}. However, these systems are currently brittle with ~\citet{noisyrag} finding that the addition of irrelevant content before a gold label answer as context to a QA system can improve performance against any intuition, suggesting much work can be done in this area to optimise how we present a model with ICL examples.

\subsection{Combined Utility of ICL Examples}
There exists a fundamental difference between \textit{relevance of documents} in IR and \textit{usefulness of examples} in ICL. In IR, a document's relevance is independent of the relevance of another document, and when combined, the information still remains relevant. The situation is more complex for ICL.
More precisely speaking, two labeled instances in ICL that are useful examples by themselves (i.e., when used as a 1-shot demonstration results in a correct prediction) may not be yielding a correct prediction when combined for a 2-shot inference \cite{lu-etal-2022-fantastically}. This is likely to happen because the decoder, on obtaining too much of a context, can be biased towards a specific topical cluster of words corresponding to the incorrect class descriptor.

While more investigation is required to analyse the empirical likelihood of this phenomenon of `non-cooperation' between examples occurring, it is worth exploring what adjustments may be needed at a methodology level to even define an ideal ranking of the training data examples for few-shot ICL. The objective in this case is not just to maximise the number of `relevant documents' (as per the IR analogy) within the top-$k$, but rather to ensure the combined usefulness of the examples. A possible direction towards this could be to adapt a listwise ranking model with this modified notion of combined relevance (usefulness).

A more computationally efficient approach would be to operate at the level of pairs, i.e., predict which pairs are concordant and discordant. An element of each pair takes on a Boolean value (either useful as a 1-shot example or not), which means that the number of different ways in which a pair can be either concordant or discordant is the number of possible Boolean functions of $2$ variables, which is $16$ (two such sample functions are Boolean OR, where if one of the examples is useful - so is the combination, and XNOR where a pair is discordant if either of the examples are useful as 1-shot).
Since, in the general case, the number of Boolean functions of $n$ variables is $2^{2^n}$, listwise training with $n>3$ will likely be computationally prohibitive.

\para{Open research questions}
Before concluding this section, we now summarise the importance of the following IR-specific research questions for ranking examples in ICL.
\uls
\li \textbf{RQ-\thesection .1}: Is ICL sensitive to the choice of a neural retrieval model, i.e., can we get an improvement using a basic Siamese model over SBERT as envisaged in \cite{rubin-etal-2022-learning}? 
\li \textbf{RQ-\thesection .2}: How faithful is the assumption that a combination of several 1-shot useful examples remain useful for ICL prediction? 
\li \textbf{RQ-\thesection .3}: If the answer to \textbf{RQ-\thesection .2} is negative, then there is a significant scope of improving over standard learning to rank approach by explicitly modeling concordance (or the lack of it) of the usefulness of examples in ICL. How can we adjust ranking models, and how much improvement can we achieve over a baseline of the standard few-shot?
\ule

\section{Informative Examples $\mapsto$ Faceted IR?} \label{sec:dicl}

In this section, we discuss the last of our proposed verticals towards an effective ICL workflow as outlined in Figure \ref{fig:icl-workflow}, which is that of seeking to provide relevant but diverse contexts to an LLM. More precisely speaking, topical diversity of the examples should play an important role in preventing a decoder bias towards a single topic. This is more true for text generation tasks, such as non-factoid question answering, where an LLM decoder needs to be aware of the different sub-topics to be able to construct a comprehensive answer.
Even for classification tasks, diverse examples are likely to help a decoder consider a majority of likely topics (the verbalisers of which map to descriptors of closely related categories) during inference, thus minimising the risks of misclassification. 

Faceted search has been well studied in IR. Explained simply, a faceted search system extracts the multiple different aspects of the information need from the top-retrieved set and maps each retrieved document to one of these aspects \cite{clarke-2009-overview,DBLP:journals/ir/GangulyJ18,DBLP:conf/sigir/GangulyLJ13,topicvis}. Faceted search is particularly useful for queries with broader information needs, where it can assist a user to reformulate their information need to one of the more specific aspects of the broader one, e.g., transform a query `dehumidifiers' to `price range of dehumidifiers' where the intention (information need facet) is to buy one \cite{CarteretteKHC14}.

Faceted search is closely related to the concept of diversified ranking \cite{alpha-ndcg}, where search systems seek to improve the retrieval effectiveness for all possible aspects of a broader information need, e.g., for the earlier example query on `dehumidifiers', retrieve documents related to information on both the aspects of price ranges, technical specifications, product reviews, and general knowledge on dehumidifiers. \citet{xquad} propose to leverage query variants (what the paper calls `sub-queries') and their top-retrieved lists for constructing a list of documents potentially relevant to each facet of the original query. Close to diversity is the concept of fair search which seeks to mitigate biases towards any particular aspects of information need, and recently neural approaches have become common to balance relevance with fairness \cite{Harrie-pl}. 

From a search user's perspective, it has been shown that diversified retrieval systems
play an important role in improving the search experience, by providing greater coverage of a topic and mitigating potential bias in search results \cite{DBLP:journals/ir/MaxwellAM19}. Similarly, a greater topical coverage and a less topical bias can potentially lead an LLM decoder towards contexts more useful for a downstream task. In fact, \citet{levy2023diverse} show that diversifying the few-shot examples on the basis of abstract syntax tree (AST) structures improves the downstream task of compositional generalisation. This indeed shows a positive direction of research where the considerable volume of work conducted on faceted search and diversification by the IR community can be useful for ICL.

However, similar to relevance, the notion of diversity would also need suitable adjustments for ICL. A suitable notion of diversity should not just consider similarities between the input examples but rather also their class labels and, more importantly, similarities in the ways in which they affect an LLM decoder's generation path. Two examples which both output similar output trees should not be considered diverse. In principle, one can potentially adapt the classification methodology that we proposed to learn the optimal number of examples based on minimising the prediction uncertainties for the purpose of classifying if a given pair of examples is diverse or not. 
Furthermore, we posit that neural approaches that take into account both relevance and fairness or diversity (again both in the context of downstream ICL) should find a use-case in ICL to help diversify the useful examples.

\para{Open research questions}
Based on the discussions in this section, we now outline the following research directions.
\uls
\li \textbf{RQ-\thesection .1}: How sensitive is ICL towards the topical diversity of the examples?
\li \textbf{RQ-\thesection .2}: How can the standard notion of diversity be extended to consider the latent dependence between the input and the output of an LLM decoder aligning towards a specific downstream task?
\li \textbf{RQ-\thesection .3}: How may existing IR metrics for diversity (e.g., $\alpha$-nDCG \cite{alpha-ndcg}) be adapted to measure how effective is the example retrieval for downstream ICL?
\li \textbf{RQ-\thesection .4}: How can multi-objective neural ranking models be trained to jointly learn downstream specific usefulness and diversity for ICL?
\ule

\section{Preliminary Evaluation} \label{sec:eval}

In this section, we report the results of our initial investigation, which was conducted to answer a subset of research questions of the first vertical, i.e., to develop an effective adaptive version of ICL that can dynamically select the number of examples.

\subsection{Research Questions and Dataset}

\para{Research Questions Investigated}
In Section \ref{ss:unsup-rankcut}, we discussed the possibilities of applying QPP-inspired unsupervised approaches for selecting a cutoff point in the ranked list of examples. On the other hand, in Section \ref{ss:sup-rankcut} we proposed a classifier-based approach to learn the optimal number of examples. In our experiments, we compare the supervised approach of Algorithm \ref{algo:buildgt} and an NQC-based unsupervised approach for adaptive $k$-shot and compare both with static $k$-shot on standard datasets for text classification.
Explicitly stated, we investigate the following research question.
\uls
\li \textbf{CRQ-1}: Does adaptively selecting the number of examples in ICL lead to improved downstream effectiveness?
\li \textbf{CRQ-2}: Does an unsupervised approach obtain a reasonable performance as compared to a supervised one? 
\ule
Since our experiments answer the above questions, they are not open, unlike the ones we expounded on in this paper. Therefore, we prefix these questions with a `C' (closed).

\para{Dataset}
We conduct experiments on three text classification datasets, namely AGNews \cite{agnews}, Jigsaw Toxic Comment\footnote{\url{https://www.kaggle.com/c/jigsaw-toxic-comment-classification-challenge}} and SST2 \cite{sst2}. Below, we provide more details on each dataset.

\uls
\li \textbf{AGNews}: AGNews is a topic classification dataset constituting news articles from the Web. Each document in the dataset belongs to one of the following 4 classes: \texttt{World}, \texttt{Sports}, \texttt{Business}, and \texttt{Sci/Tech}. The total number of training instances is $120,000$, while the test set size is $7,600$. Each class contains $30,000$ samples from the train set and $1,900$ instances from the test set.

\li \textbf{Jigsaw Toxic Comments}:
Due to its societal impact, toxicity prediction is a problem of considerable practical interest.
This dataset, released by Jigsaw and Google as a part of a Kaggle competition, comprises of comments extracted from Wikipedia's talk page, each being annotated by human evaluators across six categories representing toxic behaviors: \texttt{toxic}, `\texttt{severe toxic}', \texttt{obscene}, \texttt{threat}, \texttt{insult}, and `\texttt{identity hate}'.

\li \textbf{SST2}:
The Stanford Sentiment Treebank (SST) is a corpus with fully labeled parse trees that allows for a complete analysis of the compositional effects of sentiment in a language. The corpus consists of $11,855$ sentences extracted from movie reviews. Being parsed with the Stanford parser, it constitutes a total of $215,154$ unique phrases from the parse trees, each annotated by 3 human judges. The SST2 (also called SST-binary) dataset is a subset of SST, specifically prepared for the task of binary classification. More precisely, neutral sentences from SST were discarded, and two level, each for the negative and the positive classes were merged thus yielding two classes in total.
\ule

\subsection{Methods and Parameters}

\para{Our proposed methods for Adaptive ICL (AICL)}
As novel methods for adaptive ICL, we employ the following:
\uls
\li The supervised strategy of Algorithm \ref{algo:buildgt}, which we call supervised adaptive ICL (\textbf{SAICL}).

\li A QPP-based unsupervised strategy (as per the generic direction outlined in Section \ref{ss:unsup-rankcut}), where we compute the rank cutoff in a relatively simple way, stated as follows. First, given a top-$M$ set of candidate examples, we compute a normalised value of the NQC estimator \cite{kurland_tois12} (we employ a max normalisation, the normalisation constant being the max NQC value from the training set). 
We then quantise the normalised values into $M$ equi-spaced intervals ranging from $0$ to the max NQC value. As per the hypothesis that a higher NQC value indicates a better retrieval quality, we employ the inverse linear relation and end up selecting a value close to $0$ for higher NQC, and a value close to $M$ for smaller ones.
We call this method \textbf{QPP-AICL}.

\ule

\para{Baselines}
As baselines to compare SAICL and QPP-AICL against, we employ the following:

\uls
\item \textbf{0-shot}: This approach simply inputs an instruction without supplying any examples.
\item \textbf{Static ICL (SICL)}: This refers to the standard method of supplying a fixed number of semantically-similar examples as input, similar to \cite{liu-etal-2022-makes}. This is different from AICL in that the number of examples in the prompt is always fixed, however, the examples themselves vary for different test inputs based on semantic similarity. For a fair comparison with AICL methods, we report the results obtained with three different values of $k$: $1$, $\lceil\frac{M}{2}\rceil$ and $M$ representing the most conservative (in terms of the input size), average, and least conservative situations. In our case, $M=5$, which means that our standard ICL experiments operate with the 1-shot, 3-shot and 5-shot settings.

\ule

\para{Model and hyper-parameter settings}
Among a relatively large number of available choices for available LLMs -- either open-source models or black-box cloud APIs -- we, in particular, conduct our experiments on GPT-J \cite{su2023roformer}.
GPT-J is an open-source GPT-3-like model trained on the Pile dataset \cite{pile}. GPT-J-6B yields performance comparable to the 6.7 billion parameter GPT-3 (Curie) on a variety of tasks \cite{mesh-transformer-jax}. The maximum context length (in terms of number of tokens) of GPT-J is 2048.

In our experiments, we vary $M$ - the maximum number of examples, from 1 to 5 (for static ICL this is denoted by $k$). For a fair comparison, we use the identical prompt template (as shown in Algorithm \ref{algo:getpreds}) and greedy decoding with the same verbalizers
across all methods employed in our experiments.

\subsection{Results}

\begin{table}[t]
\centering
\caption{Macro-averaged precision, recall and F1-scores for different in-context learning (ICL) methodologies. The column $k$ denotes the number of few-shot examples. For AICL approaches, this column denotes the average number of examples used for the respective method. `AIS' denotes the average input size measured in terms of the number of tokens rounded off to the nearest integer. 
\label{table:results}
}
\begin{adjustbox}{width=0.9\columnwidth}
\begin{tabular}{llrrrrr}
\toprule
\multirow{2}{*}{} & & \multicolumn{5}{c}{Evaluation} \\ \cmidrule(l){4-7}
Dataset & Method & $k$ & Precision & Recall & F-score & AIS \\
\midrule
\multirow{6}{*}{AGNews} & 0-shot & 0 & 0.6569 & 0.5932 & 0.5849 & 60 \\
 & SICL & 1 & 0.9015 & 0.9017 & 0.9016 & 125 \\
 & SICL & 3 & 0.9008 & 0.8997 & 0.8989 & 252 \\
 & SICL & 5 & 0.8963 & 0.8930 & 0.8917 & 380 \\
 & QPP-AICL & 3 & 0.8545 & 0.8499 & 0.8486 & 220 \\
 & SAICL & 1.87 & \textbf{0.9080} & \textbf{0.9096} & \textbf{0.9067} & 175 \\
\midrule
\multirow{6}{*}{Toxicity} & 0-shot & 0 & 0.5689 & 0.6238 & 0.5769 & 103 \\
 & SICL & 1 & 0.5760 & 0.6989 & 0.5505 & 195 \\
 & SICL & 3 & 0.6092 & 0.7180 & 0.6254 & 335 \\
 & SICL & 5 & 0.6078 & \textbf{0.7248} & 0.6217 & 431 \\
 & QPP-AICL & 3 & 0.5906 & 0.6942 & 0.5977 & 289 \\
 & SAICL & 3.46 & \textbf{0.6194} & 0.6983 & \textbf{0.6303} & 359 \\
\midrule
\multirow{6}{*}{SST2} & 0-shot & 0 & 0.7503 & 0.5022 & 0.3379 & 30 \\
 & SICL & 1 & 0.8703 & 0.8703 & 0.8703 & 61 \\
 & SICL & 3 & 0.9140 & 0.9137 & 0.9137 & 121 \\
 & SICL & 5 & 0.9245 & 0.9230 & 0.9230 & 181 \\
 & QPP-AICL & 3 & 0.8556 & 0.8479 & 0.8470 & 106 \\
 & SAICL & 4.12 & \textbf{0.9302} & \textbf{0.9304} & \textbf{0.9302} & 154 \\
\bottomrule
\end{tabular}
\end{adjustbox}
\end{table}

Table \ref{table:results} shows the results (in terms of macro-averaged precision, recall and F1)  obtained by the different ICL strategies.
It can be seen that SAICL turns out to be the best among the competing approaches. The reason it outperforms the best baseline (static ICL) is that SAICL is able to effectively adapt the number of examples to use, thereby preventing itself from the degradation effects of non-relevant (not useful) examples.
In effect, it learns a latent relationship between the topical content and the quantity of context required to guide the decoder's output in the right direction effectively.  %
Moreover, SAICL is able to operate more effectively with smaller input sizes (see the average value of $k$ and also the average size of the input in terms of the number of tokens), which means that it is computationally faster as compared to static ICL (SICL).
Our observations reveal that \textbf{CRQ-1} is answered in the affirmative, i.e., an adaptive selection of the number of examples in ICL does improve downstream effectiveness and efficiency.

The results with the unsupervised QPP-based approach (QPP-AICL)
turned out to be worse than the baseline of static ICL. From a broader perspective, this points to an important finding - that off-the-shelf IR approaches without modifications specifically suited to the underlying characteristics of the downstream tasks in ICL may not directly yield improvements in the effectiveness of ICL. For instance, NQC seeks to estimate relevance of documents, and as we have argued before, that relevance has a different interpretation for the ICL examples.
Although the observations with QPP-AICL answers \textbf{CRQ-2} in negative, i.e., an unsupervised approach for an adaptive selection of ICL examples is substantially worse than a supervised one, they do 
suggest that methodologies developed by researchers in the future for answering any of the open research questions discussed in this paper should be fundamentally grounded in modeling the notion of relevance (usefulness of examples) in a robust and effective manner.

\section{Conclusion} \label{sec:conclusion}

In this perspective paper, we discuss how some of the recent developments in generative AI (specifically in-context learning or ICL) can provide a scope to IR/NLP researchers to revisit some of the well-researched IR topics in a new light, where the notion of relevance of a document to an information need changes to that of usefulness of a few-shot example for a downstream AI task, e.g., text classification, question answering etc. More specifically, we suggest three main verticals in which this research can be structured - each offering a set of open questions related to core IR research.

The first vertical aims at adaptively adjusting an ICL workflow, e.g., choosing the number of examples to be used in a data-driven manner. Initial empirical investigations reported in this perspective paper shows that this direction is promising. The second vertical mainly covers devising novel ranking models to better distinguish (and thereby retrieve at better ranks) a useful few-shot context from a noisy one. Finally, the third vertical concerns an investigation of topical diversity in the few-shot examples for better downstream prediction.

We believe that the research questions that we have proposed in this paper will benefit the research community to exploit this synergy between ICL and IR, and eventually guide the development of new algorithms and techniques.

\bibliographystyle{ACM-Reference-Format}
\bibliography{refs}

\end{document}